\documentclass[twocolumn,aps,prc,superscriptaddress,showpacs,floatfix,longbibliography,nofootinbib]{revtex4-1}
\usepackage{url}
\usepackage{cancel}
\usepackage[colorlinks,linkcolor=blue,citecolor=blue,filecolor=black,urlcolor=blue]{hyperref}
\usepackage{epsfig,graphics}
\usepackage{graphicx}
\usepackage{dcolumn}
\usepackage{bm}
\usepackage[usenames]{color}
\usepackage{amssymb}
\usepackage{amsmath}
\usepackage{multirow}
\usepackage{float}
\usepackage{harpoon}
\usepackage{MnSymbol}
\usepackage{appendix}
\usepackage{color}
\usepackage{hyperref}
\usepackage{cleveref}
\usepackage{dutchcal}
\usepackage[normalem]{ulem}

\begin{document}

\title{Symmetry energy of baryon- and neutron-rich nuclear matter}
\author{Zhi-Ying Qin}
\affiliation{School of Physics Science and Engineering, Tongji University, Shanghai 200092, China}
\author{Jia Zhou}
\affiliation{School of Physics Science and Engineering, Tongji University, Shanghai 200092, China}
\author{Jun Xu}\email{junxu@tongji.edu.cn}
\affiliation{School of Physics Science and Engineering, Tongji University, Shanghai 200092, China}
\affiliation{Southern Center for Nuclear-Science Theory (SCNT), Institute of Modern Physics, Chinese Academy of Sciences, Huizhou 516000, Guangdong Province, China}
\begin{abstract}
Based on the relativistic mean-field model and assuming $G$-parity invariance, we have studied the equation of state of baryon- and neutron-rich matter produced in low-energy relativistic heavy-ion collisions. Similar to the traditional isospin symmetry energy, we define the baryon-antibaryon symmetry energy characterizing the energy difference due to the baryon-antibaryon asymmetry. The potential difference between nucleons and antinucleons is correlated with the potential contribution of the baryon-antibaryon symmetry energy mainly from the vector interaction in baryon-rich matter. The isospin symmetry energy is considerably reduced even with a small fraction of antinucleons compared to the traditional case with only nucleons. A more attractive antineutron potential than antiproton potential is observed, and the isospin splitting of the mean-field potential for antinucleons is found to be intrinsically larger than that for nucleons in baryon- and neutron-rich matter.
\end{abstract}
\maketitle


Understanding accurately the nuclear matter equation of state (EOS) is one of the fundamental goals of nuclear physics. The traditional nuclear symmetry energy $E_{sym}$, which describes the energy difference between isospin symmetric nuclear matter and neutron-rich nuclear matter, is the most uncertain part of the nuclear matter EOS, and it has important ramifications in nuclear structures, nuclear reactions, and nuclear astrophysics~\cite{Steiner:2004fi,Steiner:2004fi,Baran:2004ih,Li:2008gp,Baldo:2016jhp,Oertel:2016bki,Tsang:2011ju,Horowitz:2014bja}. While the kinetic contribution of $E_{sym}$ originates from different Fermi surfaces of neutrons and protons in neutron-rich matter, the potential contribution of $E_{sym}$ originates from different neutron-proton and neutron-neutron interactions, correlated with the different mean-field potentials of neutrons and protons in neutron-rich nuclear matter.

In general, any energy difference due to the asymmetry of nuclear matter can be used to define the symmetry energy. For instance, the spin symmetry energy characterizes the energy excess due to the spin polarization of nuclear matter~\cite{Tachibana:2025wey,Khoa:2022yee,Chamel:2010wr,Vidana:2002pc}, and it is related to the spin-dependent nuclear interaction. In relativistic heavy-ion collisions, antinucleons are produced, and the symmetry energy can be generalized to that describing the energy difference due to the asymmetry of nuclear matter and antinuclear matter, correlated with the different nucleon and antinucleon mean-field potentials in baryon-rich nuclear matter. According to the $G$-parity invariance, the vector potential changes sign for antinucleons in nuclear matter, while the scalar potential remains the same. Similar procedures have been applied to studies on baryon-rich quark matter, and it has been found that the strength of the vector potential is important in determining the EOS as well as the phase diagram of baryon-rich strong-interacting matter~\cite{Xu:2013sta,Carignano:2010ac,Bratovic:2012qs,Costa:2015bza}. In addition, different degrees of freedom (DOF) often interact with each other. For example, the isospin symmetry energy $E_{sym}$ can be modified in the presence of the strangeness DOF~\cite{Yang:2025wop,Providencia:2012rx,Bednarek:2019xyt}. It is of interest to study how the traditional isospin symmetry energy $E_{sym}$ will be affected by the occurrence of antinucleons.


The purpose of the present study is to discuss the symmetry energy of baryon- and neutron-rich nuclear matter, which can be produced in beam-energy-scan or fixed-target experiments at relativistic heavy-ion collider (RHIC), based on the relativistic mean-field model (RMF). We start from the following Lagrangian
\begin{eqnarray}
\mathcal{L} = \mathcal{L}_{nm} + \mathcal{L}_{\sigma} + \mathcal{L}_{\omega} + \mathcal{L}_{\rho} + \mathcal{L}_{\delta} + \mathcal{L}_{\omega\rho},
\end{eqnarray}
with
\begin{eqnarray}
		\mathcal{L}_{nm} &=& \bar{\psi}(i\gamma^\mu\partial_\mu - m)\psi + g_\sigma\sigma\bar{\psi}\psi - g_\omega\bar{\psi}\gamma^\mu\omega_\mu\psi \notag\\
		&-& \frac{g_\rho}{2}\bar{\psi}\gamma^\mu\bm{\rho}_\mu\bm{\tau}\psi + g_\delta \bar{\psi}\bm{\delta}\bm{\tau}\psi,\\
		\mathcal{L}_\sigma &=& \frac{1}{2}(\partial^\mu\sigma\partial_\mu\sigma - m_\sigma^2\sigma^2) - \frac{A}{3}\sigma^3 - \frac{B}{4}\sigma^4,\\
		\mathcal{L}_\omega &=& -\frac{1}{4}F^{\mu\nu}F_{\mu\nu} + \frac{1}{2}m_\omega^2\omega_\mu\omega^\mu + \frac{C}{4}(g_\omega^2\omega_\mu\omega^\mu)^2,\\
		\mathcal{L}_\rho &=& -\frac{1}{4}\bm{B}^{\mu\nu}\bm{B}_{\mu\nu} + \frac{1}{2}m_\rho^2\bm{\rho}_\mu\bm{\rho}^\mu,\\
		\mathcal{L}_\delta &=& \frac{1}{2}(\partial^\mu\bm{\delta}\partial_\mu\bm{\delta} - m_\delta^2\bm{\delta}^2),\\
		\mathcal{L}_{\omega\rho} &=& \frac{1}{2}\alpha'_3g_\omega^2g_\rho^2\omega_\mu\omega^\mu\bm{\rho}_\mu\bm{\rho}^\mu.
\end{eqnarray}
In the above, $\mathcal{L}_{nm}$ represents the kinetic part of nucleons and antinucleons as well as the coupling to other mesons, with $\psi$, $\sigma$, $\omega_\mu$, $\bm{\rho}$, and $\bm{\delta}$ being the fields of nucleons as well as antinucleons and corresponding mesons, $g_\sigma$, $g_\omega$, $g_\rho$, and $g_\delta$ being the corresponding coupling coefficients, and $\bm{\tau}$ being the Pauli matrices in isospin space. $\mathcal{L}_\sigma$, $\mathcal{L}_\omega$, $\mathcal{L}_\rho$, and $\mathcal{L}_\delta$ stand for the free and self-interacting terms of $\sigma$, $\omega$, $\rho$, and $\delta$ mesons, and $\mathcal{L}_{\omega\rho}$ is the cross interaction contribution between $\omega$ and $\rho$ mesons. The antisymmetric field tensors are defined as $F_{\mu\nu} = \partial_\nu \omega_\mu - \partial_\mu \omega_\nu$ and $\bm{B}_{\mu\nu} = \partial_\nu \bm{\rho}_\mu - \partial_\mu \bm{\rho}_\nu - g_\rho (\bm{\rho}_\mu \times \bm{\rho}_\nu)$. Values of the coefficients are chosen to be $g_\sigma^2/m_\sigma^2=2.1853\times10^{-4}$ MeV$^{-2}$, $g_\omega^2/m_\omega^2=1.1255\times10^{-4}$ MeV$^{-2}$, $g_\rho^2/m_\rho^2=2.3149\times10^{-4}$ MeV$^{-2}$, $g_\delta^2/m_\delta^2=2.6175\times10^{-5}$ MeV$^{-2}$, $A=397.63$ MeV, $B=122.41$, $C=0.048880$, and $\alpha'_3=0.051784$, and they lead to the binding energy $-16$ MeV of normal nuclear matter, the symmetry energy $E_{sym}^0=30$ MeV as well as its slope parameter $L=60$ MeV, the incompressibility $K_0=230$ MeV as well as the skewness parameter $Q_0=-995$ MeV, the isoscalar nucleon effective mass $m_s^\star/m=0.80$, and the isovector nucleon effective mass $m_v^\star/m=0.83$, at the saturation density $\rho_0=0.16$ fm$^{-3}$. We have also used $Q_0^\prime=1835$ MeV with the same values of other parameters in order to investigate the effect of the vector potential.

In the mean-field approximation, the meson fields in the static baryon- and neutron-rich nuclear matter can be calculated through the following equations obtained from the Euler-Lagrange equation
\begin{eqnarray}
g_\sigma\rho_s&=&m_\sigma^2\left\langle \sigma\right\rangle+ A\left\langle \sigma\right\rangle^2 + B\left\langle \sigma\right\rangle^3, \label{rhos}\\
g_\omega\rho&=&m_\omega^2\left\langle \omega_0 \right\rangle+ Cg_\omega(g_\omega\left\langle \omega_0 \right\rangle)^3+\alpha'_3g_\omega^2g_\rho^2\left\langle \omega_0 \right\rangle\left\langle \rho_0 \right\rangle^2, \label{rho}\\
\frac{g_\rho}{2}\rho_{3}&=&m_\rho^2\left\langle \rho_0 \right\rangle+\alpha'_3g_\omega^2g_\rho^2\left\langle \omega_0 \right\rangle^2\left\langle \rho_0 \right\rangle, \label{rho3}\\
		g_\delta\rho_{s3}&=&m_\delta^2\left\langle \delta\right\rangle, \label{rhos3}
\end{eqnarray}
where the isoscalar and isovector components of the scalar and vector density are
\begin{eqnarray}
&&		\rho_s=\rho_{sp}+\rho_{sn},~\rho_{s3}=\rho_{sp}-\rho_{sn}, \notag\\
&&		\rho=\rho_B-\rho_{\bar{B}}=(\rho_{p}+\rho_{n})-(\rho_{\bar{p}}+\rho_{\bar{n}}), \notag\\
&& \rho_{3}=(\rho_{p}-\rho_{\bar{p}})-(\rho_{n}-\rho_{\bar{n}}). \notag
\end{eqnarray}
The above densities can be calculated from
\begin{eqnarray}
\rho_{sq}&=&2\int\frac{d^3k}{\left(2\pi\hbar\right)^3}\frac{M_{q}^\star}{\sqrt{k^2+{M_q^\star}^2}} [f_q(k)+f_{\bar{q}}(k)],\\
\rho_{q,\bar{q}}&=&2\int\frac{d^3k}{\left(2\pi\hbar\right)^3} f_{q,\bar{q}}(k),
\end{eqnarray}
where $q=n,p$ is the isospin index,
\begin{eqnarray}
M_q^\star = m-g_\sigma\left\langle \sigma\right\rangle \pm g_\delta\left\langle \delta\right\rangle
\end{eqnarray}
is the Dirac mass, with the `$+$' sign for $q=n$ and the `$-$' sign for $q=p$, and
\begin{eqnarray}
f_{q,\bar{q}}(k)=\frac{1}{1+e^{(\sqrt{k^2+{M_q^\star}^2} - \tilde{\mu}_{q,\bar{q}})/T}}
\end{eqnarray}
is the phase-space distribution function, where the effective chemical potential is defined as
\begin{eqnarray}\label{mu}
\tilde{\mu}_{n,p}=\mu_{n,p}-g_\omega\left\langle \omega_0 \right\rangle \pm \frac{g_\rho}{2}\left\langle \rho_0 \right\rangle,
\end{eqnarray}
with the `$+$' sign for $q=n$ and the `$-$' sign for $q=p$. In the chemical-equilibrium case, the relation $\tilde{\mu}_{\bar{n},\bar{p}}=-\tilde{\mu}_{n,p}$ is valid. In the more general case without the constraint of chemical equilibrium, $\mu_{q,\bar{q}}$ can be determined individually by the density $\rho_{q,\bar{q}}$ for each species $q$ at the temperature $T$. The chemical-equilibrium case is closer to the situation in relativistic heavy-ion collisions, while the chemical-nonequilibrium case is more clear in the theoretical point of view.

The energy density of the nucleon-antinucleon system can be expressed as
\begin{eqnarray}
\mathcal{E} = \mathcal{E}_p + \mathcal{E}_k + m\rho_{total},
\end{eqnarray}
where
\begin{eqnarray}\label{Ep}
\mathcal{E}_p &=&  \frac{1}{2} m_\sigma^2 \left\langle \sigma \right\rangle ^2 + \frac{A}{3} \left\langle \sigma \right\rangle^3 + \frac{B}{4} \left\langle \sigma \right\rangle^4 - \frac{1}{2} m_\omega^2 \left\langle \omega_0\right\rangle ^2 \notag\\
&-& \frac{C}{4} (g_\omega^2 \left\langle \omega_0\right\rangle ^2)^2
- \frac{1}{2} m_\rho^2 \left\langle \rho_0\right\rangle ^2
+ g_\omega \left\langle \omega_0\right\rangle  \rho + \frac{g_\rho}{2} \left\langle \rho_0\right\rangle \rho_3 \notag\\
&+& \frac{1}{2} m_\delta^2 \left\langle \delta\right\rangle ^2
- \frac{1}{2} \alpha'_3 g_\omega^2 g_\rho^2 \left\langle \omega_0\right\rangle^2 \left\langle \rho_0\right\rangle^2 \notag\\
&+& \sum_{q} (M_q^\star - m)(\rho_q+\rho_{\bar{q}})
\end{eqnarray}
and
\begin{eqnarray}
\mathcal{E}_k = \sum_q 2\int\frac{d^3k}{\left(2\pi\hbar\right)^3} (\sqrt{k^2+{M_q^\star}^2}-M_q^\star) [f_q(k)+f_{\bar{q}}(k)] \notag\\
\end{eqnarray}
are respectively the potential and kinetic contributions, and $\rho_{total}=\sum_{q}(\rho_q+\rho_{\bar{q}})$ is the total density of all particle species. The energy per particle after subtracting the nucleon mass can then be calculated from
\begin{eqnarray}\label{E}
E=(\mathcal{E}_p + \mathcal{E}_k)/\rho_{total}=E_p+E_k,
\end{eqnarray}
which has the potential contribution $E_p$ and the kinetic contribution $E_k$. We may also calculate the Schr\"odinger equivalent potential~\cite{Jaminon:1989wj} of neutrons and protons in the present RMF framework
\begin{eqnarray}\label{SEP}
U_{q}&=&\Sigma_{sq}+\Sigma_{vq} + \frac{\sqrt{k^2+{M_q^\star}^2}+\Sigma_{vq}-m}{m}\Sigma_{vq}\notag\\
&+&\frac{1}{2m}\left( \Sigma_{sq}^2-\Sigma_{vq}^2\right),
\end{eqnarray}
where the scalar and vector self-energies are, respectively,
\begin{eqnarray}
		\Sigma_{sq}=&-g_\sigma\left\langle \sigma\right\rangle \pm g_\delta\left\langle \delta\right\rangle,\\
\Sigma_{vq}=&g_\omega\left\langle \omega_0 \right\rangle \mp \frac{g_\rho}{2}\left\langle \rho_0 \right\rangle,
\end{eqnarray}
with the upper sign for $q=n$ and the lower sign for $q=p$. For the Schr\"odinger equivalent potential $U_{\bar{q}}$ of antinucleons, the vector self-energy $\Sigma_{vq}$ in Eq.~(\ref{SEP}) changes sign.


\begin{figure}[ht]
\includegraphics[width=1\linewidth]{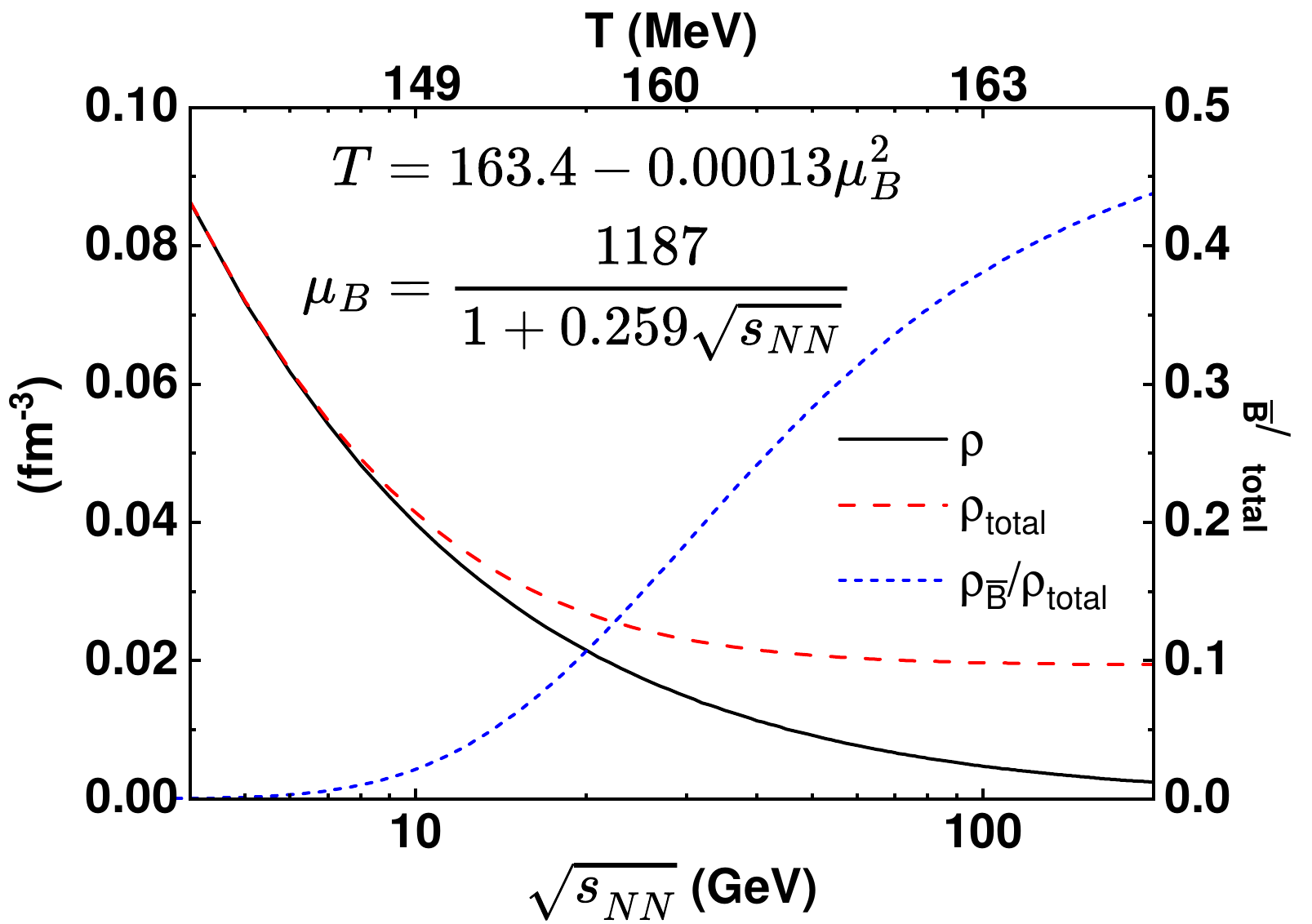}
\caption{\label{fig1} Net nucleon density $\rho$, total density $\rho_{total}$, and the fraction of antinucleons $\rho_{\bar{B}}/\rho_{total}$ at chemical freeze-out in relativistic heavy-ion collision at different energies.}
\end{figure}

\begin{figure}[ht]
\includegraphics[width=0.7\linewidth]{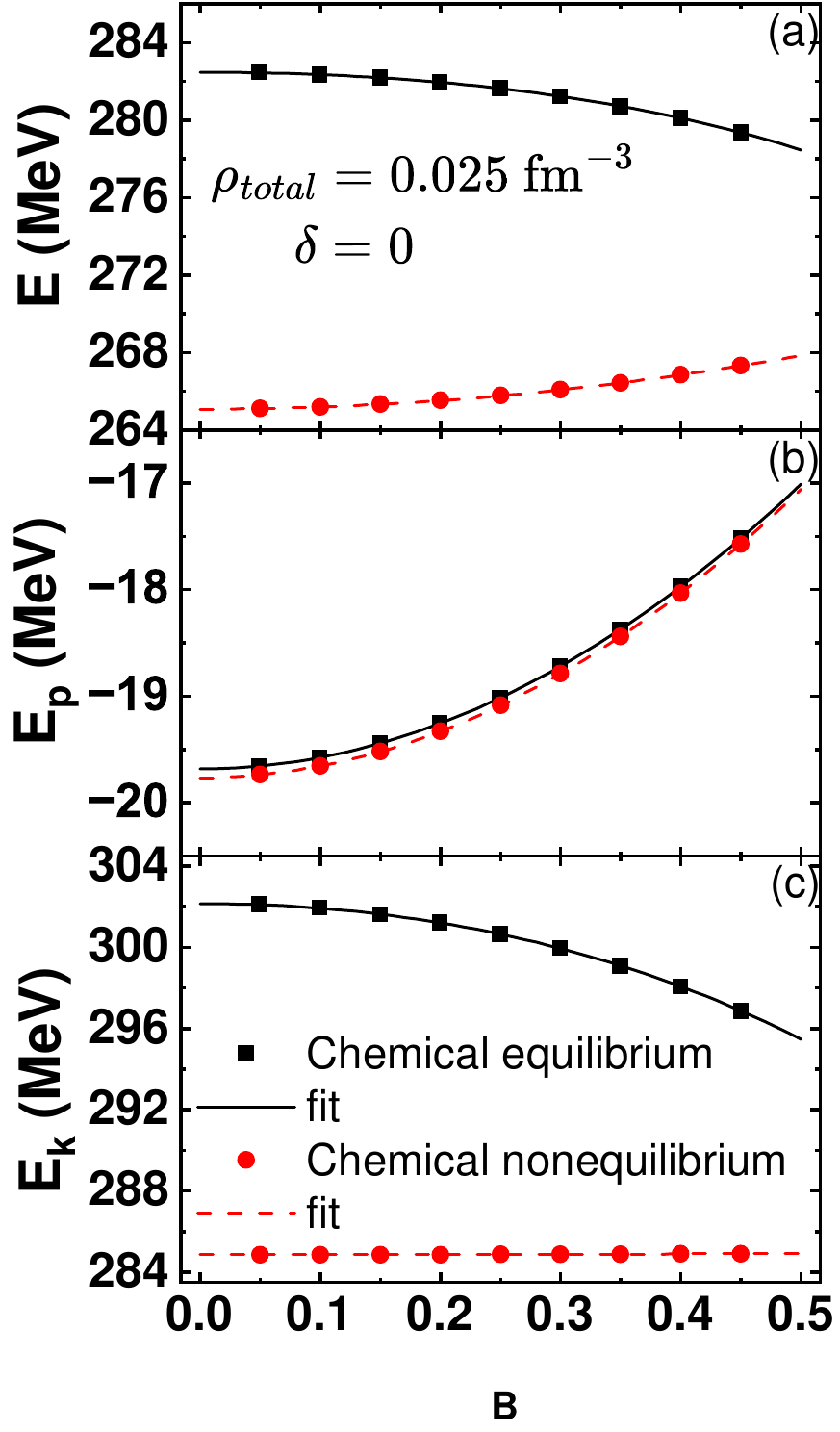}
\caption{\label{fig2} Energy per particle (a) as well as its potential energy (b) and kinetic energy (c) contribution in isospin-symmetric baryon-rich nuclear matter at different baryon-antibaryon asymmetries and a fixed total density $\rho_{total}=0.025$ fm$^{-3}$. Results from the chemical-equilibrium case at $T=165.9 \sim 168.5$ MeV and the chemical-nonequilibrium case at $T=160$ MeV are compared, and they are fitted by Eq.~(\ref{fit}).}
\end{figure}

\begin{figure}[ht]
\includegraphics[width=0.7\linewidth]{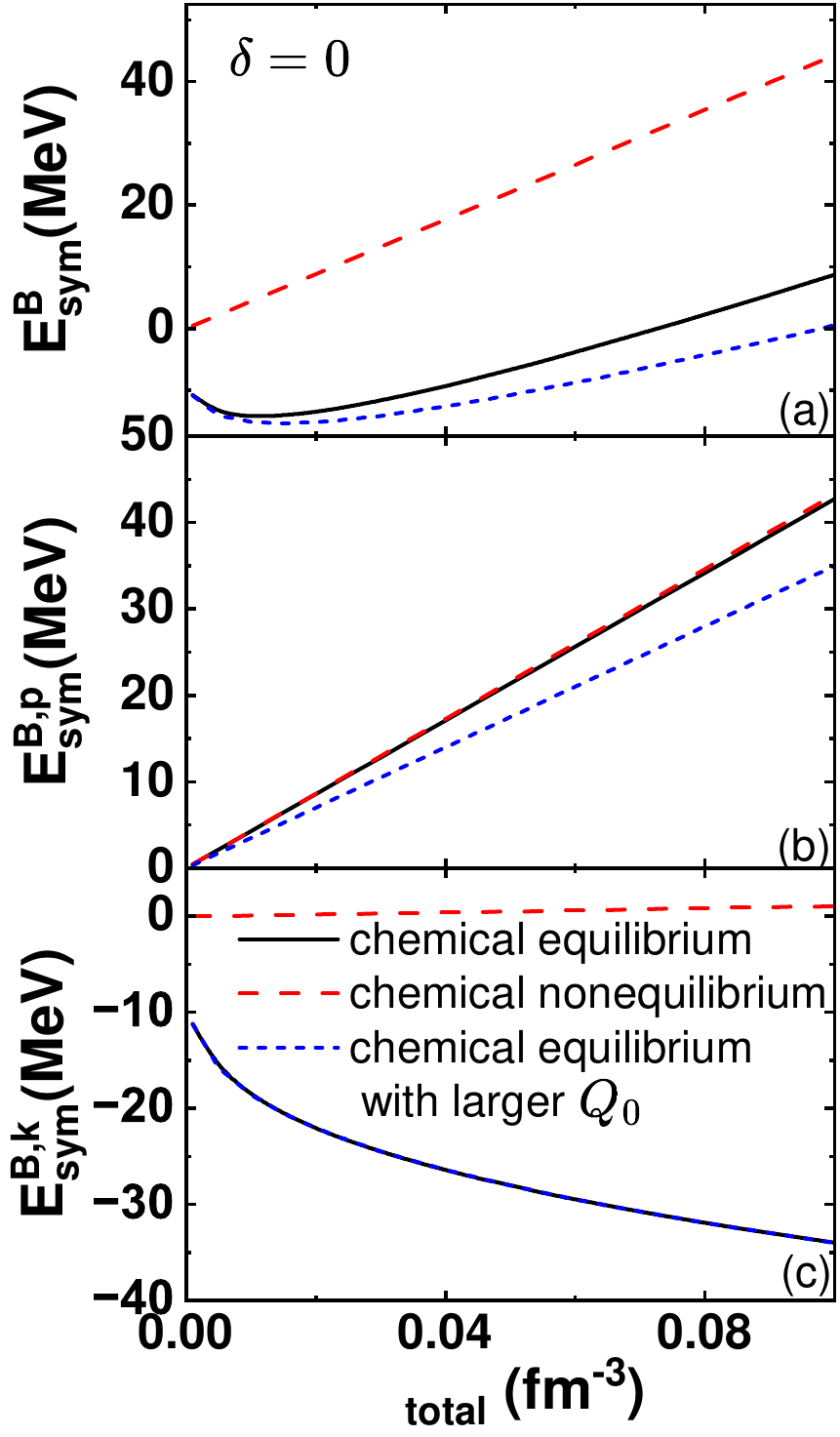}
\caption{\label{fig3} Baryon-antibaryon symmetry energy (a) as well as its potential energy (b) and kinetic energy (c) contribution as a function of total density in isospin-symmetric baryon-rich nuclear matter. Results from the chemical-equilibrium case with different vector interactions varied by $Q_0$ and the chemical-nonequilibrium case ($T=160$ MeV) are compared.}
\end{figure}

We now first illustrate the densities at chemical freeze-out in relativistic heavy-ion collisions at different energies in Fig.~\ref{fig1}, with the dependence of the temperature $T$ and the baryon chemical potential $\mu_B$ on the center-of-mass energy $\sqrt{s_{NN}^{}}$ taken from Ref.~\cite{Sharma:2026ucy}, and the nucleon and antinucleon interaction based on the above RMF model. Other baryons and antibaryons than nucleons and antinucleons are neglected here, and the system is assumed to be isospin symmetric, i.e., $\mu_B=\mu_n=\mu_p$ in Eq.~(\ref{mu}), in this illustration. It is seen that both the net nucleon density $\rho$ and the total density $\rho_{total}$ decreases with increasing $\sqrt{s_{NN}^{}}$, while the fraction of antibaryon density increases with increasing $\sqrt{s_{NN}^{}}$. The system is rather baryon-rich at low $\sqrt{s_{NN}^{}}$, while it almost reaches baryon-antibaryon symmetry at high $\sqrt{s_{NN}^{}}$. Around $\sqrt{s_{NN}^{}}=25$ GeV, the temperature at chemical freeze-out is about 160 MeV, and the total density is about $\rho_{total}\approx0.025$ fm$^{-3}$ together with an antinucleon fraction of $\rho_{\bar{B}}/\rho_{total}\approx0.14$.

We can calculate the energy per particle as well as the kinetic and potential contribution according to Eq.~(\ref{E}) in hot baryon-rich nuclear matter. Taking $\rho_{total}=0.025$ fm$^{-3}$ as an example, results of the energy per particle for the chemical-equilibrium and chemical-nonequilibrium cases are shown in Fig.~\ref{fig2}. In the chemical-equilibrium case, the baryon-antibaryon asymmetry $\delta_B = (\rho_B-\rho_{\bar{B}})/\rho_{total}$ is determined by the temperature $T$ at a fixed $\rho_{total}$. From $\delta_B=0.05$ to 0.45, the temperature decreases from 168.5 MeV to 165.9 MeV, and no solutions can be found for very small $\delta_B$ in the chemical-equilibrium case. In the chemical-nonequilibrium case, the temperature is fixed at $T=160$ MeV. It is seen that the potential energy generally increases with increasing $\delta_B$ and is similar in the two cases, where the difference mainly originates from different temperatures. On the other hand, the kinetic energy decreases with increasing $\delta_B$ due to a higher temperature at smaller $\delta_B$ in the chemical-equilibrium case, while it remains almost unchanged with increasing $\delta_B$ in the chemical-nonequilibrium case due to the smearing of the Fermi surfaces of nucleons and antinucleon at such high temperatures. Consequently, the total energy per particle decreases with increasing $\delta_B$ in the chemical-equilibrium case but increases with increasing $\delta_B$ in the chemical-nonequilibrium case. To extract the baryon-antibaryon symmetry energy
\begin{eqnarray}
E_{sym}^B = \frac{1}{2} \left(\frac{\partial^2 E}{\partial \delta_B^2}\right)_{\rho_{total}}
\end{eqnarray}
as well as the kinetic and potential contribution, we fit the $\delta_B$ dependence of the energy per particle in Fig.~\ref{fig2} according to
\begin{eqnarray}\label{fit}
E = E_0^B + E_{sym}^B \delta_B^2 + E_{sym}^{B(4)} \delta_B^4
\end{eqnarray}
indicated by the lines. The odd terms of $\delta_B$ vanish based on the present RMF framework. We found that the contribution of the $\delta_B^4$ term is larger in the chemical-equilibrium case than in the chemical-nonequilibrium case, mostly due to the kinetic contribution in the former case.

The density dependence of the baryon-antibaryon symmetry energy $E_{sym}^B$ as well as its potential and kinetic energy contribution is shown in Fig.~\ref{fig3}, obtained from the fitting method as described in Fig.~\ref{fig2}. The potential contribution $E_{sym}^{B,p}$ increases almost linearly with increasing $\rho_{total}$ and is similar in the chemical-equilibrium and chemical-nonequilibrium case, and it becomes weaker from a weaker vector interaction with a larger $Q_0$ at subsaturation densities. For the kinetic contribution $E_{sym}^{B,k}$, it is almost zero in the chemical-nonequilibrium case as already seen in Fig.~\ref{fig2}, but is negative and decreases with increasing $\rho_{total}$ in the chemical-equilibrium case. The latter originates from a more rapid decrease of $E_k$ with increasing $\delta_B$ at larger $\rho_{total}$, since a larger variation of the temperature is needed to vary $\delta_B$ at larger $\rho_{total}$. Consequently, the $E_{sym}^B$ increases almost linearly with increasing $\rho_{total}$ in the chemical-nonequilibrium case, but it first decreases then increases with increasing $\rho_{total}$ in the chemical-equilibrium case. In the latter case, the $E_{sym}^B$ can be negative due to the negative kinetic contribution, and is smaller with a weaker vector interaction.

\begin{figure}[ht]
\includegraphics[width=1\linewidth]{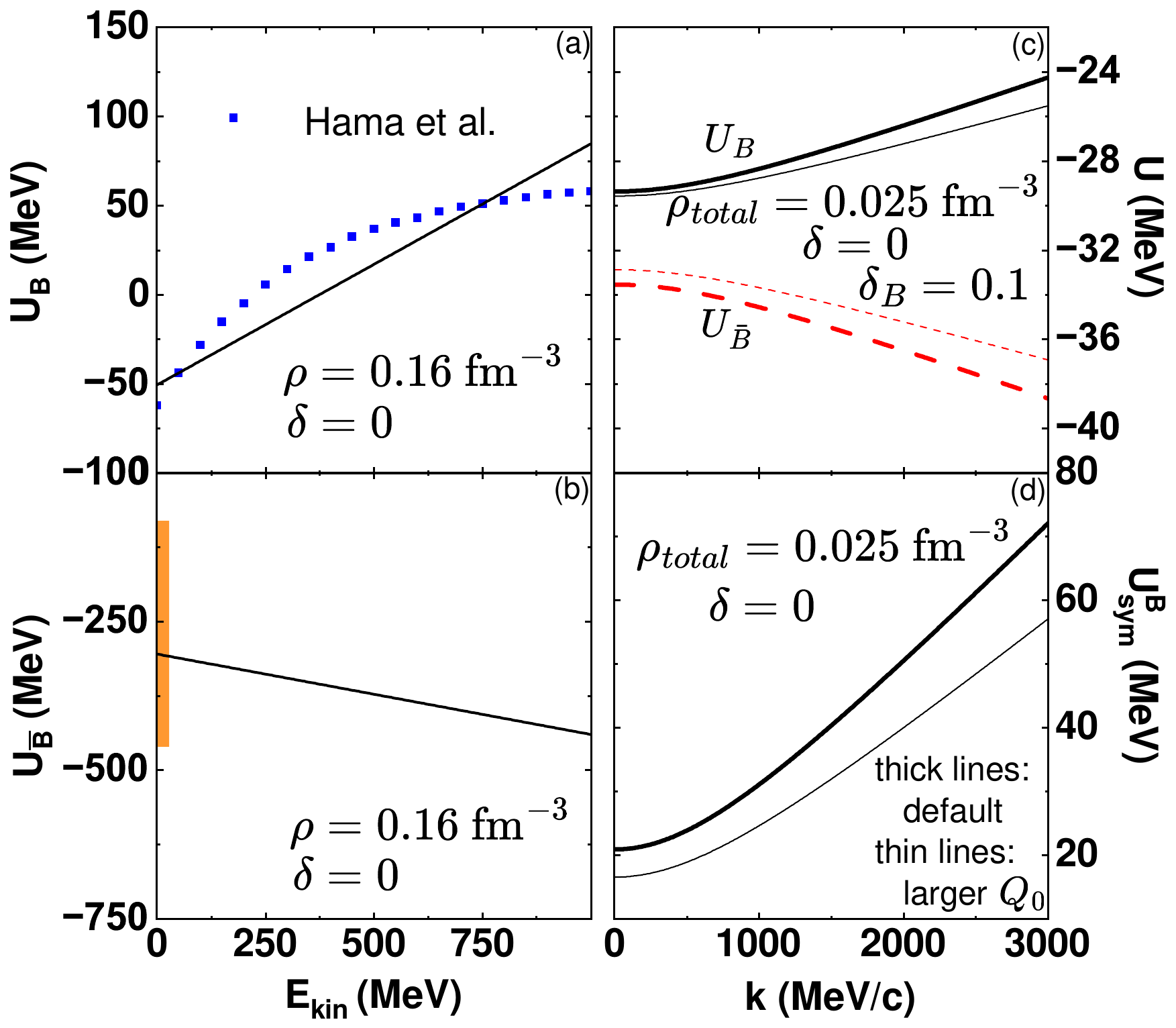}
\caption{\label{fig4} Left: Nucleon (upper) and antinucleon (lower) Schr\"odinger equivalent potential as a function of particle energy in cold normal nuclear matter; Right: Nucleon and antinucleon Schr\"odinger equivalent potential (upper) and the baryon-antibaryon symmetry potential (lower) as a function of particle momentum in hot isospin-symmetric baryon-rich nuclear matter at a fixed total density $\rho_{total}=0.025$ fm$^{-3}$. }
\end{figure}

The Schr\"odinger equivalent potentials of nucleons ($U_B$) and antinucleons ($U_{\bar{B}}$) in isospin-symmetric baryon-rich nuclear matter are displayed in Fig.~\ref{fig4}. The left panels show the dependence of nucleon and antinucleon potentials on $E_{kin}=\sqrt{k^2+{M_q^\star}^2}+\Sigma_{vq}-m$ in cold normal nuclear matter. The nucleon potential, which is attractive at small $E_{kin}$ and repulsive at large $E_{kin}$, increases linearly with increasing $E_{kin}$ as can be seen from Eq.~(\ref{SEP}), roughly consistent with the optical potential extracted from proton-nucleus scatterings by Hama et al.~\cite{Hama:1990vr,Cooper:1993nx}, as shown in Fig.~\ref{fig4} (a). The antinucleon potential, which is deeply attractive, decreases linearly with increases $E_{kin}$, and it is within the band of the antiproton potential extracted based on studies of antiprotonic atoms~\cite{Barnes:1972wds,Poth:1977kg,Batty:1981qj,Wong:1984fy} at $E_{kin}=0$, as shown in Fig.~\ref{fig4} (b). In hot baryon-rich nuclear matter at $\rho_{total}=0.025$ fm$^{-3}$ and $\delta_B=0.1$, with $T \approx 168.5$ MeV from the chemical-equilibrium condition, the momentum dependence of the nucleon and antinucleon potential is shown in Fig.~\ref{fig4} (c). Both potentials are attractive but $U_{\bar{B}}$ is more attractive than $U_B$, and their difference becomes smaller with a weaker vector interaction from a larger $Q_0$. It is seen that $U_B$ ($U_{\bar{B}}$) increases (decreases) with increasing nucleon momentum. Similar to the isospin symmetry potential, we can define the baryon-antibaryon symmetry potential in baryon-rich nuclear matter as
\begin{eqnarray}
U_{sym}^B = (U_B-U_{\bar{B}})/2\delta_B.
\end{eqnarray}
It is seen in Fig.~\ref{fig4} (d) that $U_{sym}^B$ is positive, due to the more attractive potential of antinucleons than nucleons in baryon-rich matter, and is larger at larger particle momenta. A weaker vector interaction leads to a smaller $U_{sym}^B$, correlated with the smaller $E_{sym}^{B,p}$ in Fig.~\ref{fig3} (b).

\begin{figure}[ht]
\includegraphics[width=0.7\linewidth]{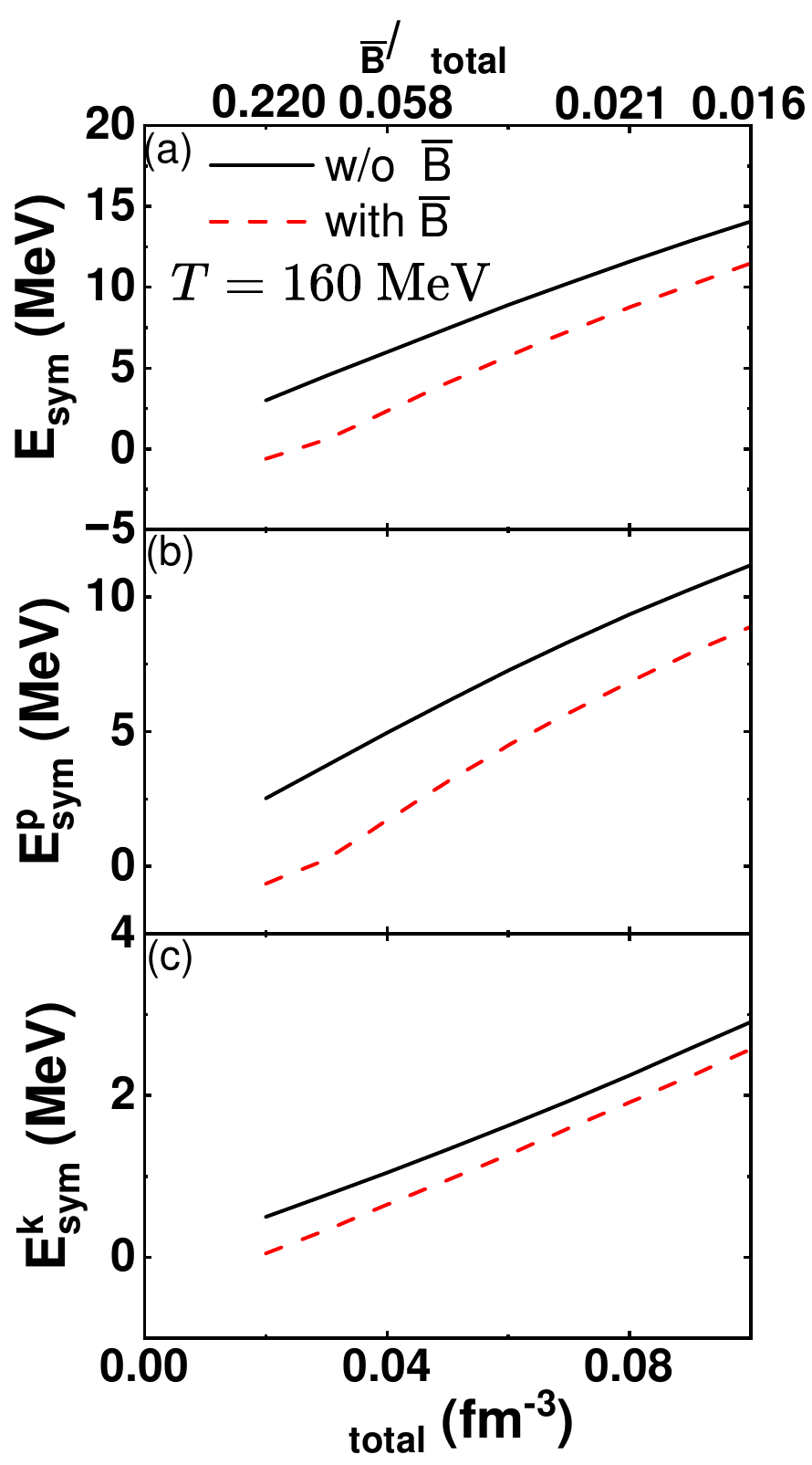}
\caption{\label{fig5} Isospin symmetry energy (a) as well as its potential energy (b) and kinetic energy contribution (c) as a function of total density in baryon-rich nuclear matter at the temperature $T=160$ MeV with and without antinucleons.}
\end{figure}

So far we have discussed the baryon-antibaryon symmetry energy and symmetry potential in isospin-symmetric baryon-rich nuclear matter. Now we move to the discussion on isospin-asymmetric baryon-rich nuclear matter. Figure~\ref{fig5} compares the isospin symmetry energy as well as its potential and kinetic contribution, which are obtained by a similar fitting as Eq.~(\ref{fit}) but with respect to the isospin asymmetry with and without antinucleons at $T=160$ MeV. At such high temperature, the total density can not be smaller than 0.02 fm$^{-3}$ in the chemical-equilibrium case assumed here. One sees that all results increase almost linearly with increasing $\rho_{total}$ based on the RMF model. The isospin symmetry energy without antinucleons is generally larger than that with antinucleons for a given $\rho_{total}$. In the latter case, we find that the baryon-antibaryon asymmetry $\delta_B$ decreases with increasing isospin asymmetry $\delta$ for fixed temperature and $\rho_{total}$, and this reduces the total potential energy through $E_{sym}^{B,p}$ especially at small $\rho_{total}$, leading to a slightly negative $E_{sym}^p$ and $E_{sym}$ there. It is noteworthy that the fraction of antinucleons $\rho_{\bar{B}}/\rho_{total}$ decreases with increasing $\rho_{total}$, and at $\rho_{total}\approx0.1$ fm$^{-3}$ the $E^p_{sym}$ is reduced by about $20\%$ even with an about $2\%$ fraction of antinucleons. This could be understood in the following way. The occurrence of antinucleons reduces $\langle \omega_0 \rangle$ and $\langle \rho_0 \rangle$ according to Eqs.~(\ref{rho}) and (\ref{rho3}), and this effect is enlarged by a factor of 4 in the potential part of the energy density according to Eq.~(\ref{Ep}) due to the fourth power of $\langle \omega_0 \rangle$ and $\langle \rho_0 \rangle$. Additional effect comes from the definition of the isospin symmetry, which is $\delta=[(\rho_n-\rho_{\bar{n}})-(\rho_p-\rho_{\bar{p}})]/\rho$ rather than $\delta=[(\rho_n-\rho_{\bar{n}})-(\rho_p-\rho_{\bar{p}})]/\rho_{total}$ in the present study.

\begin{figure}[ht]
\includegraphics[width=0.8\linewidth]{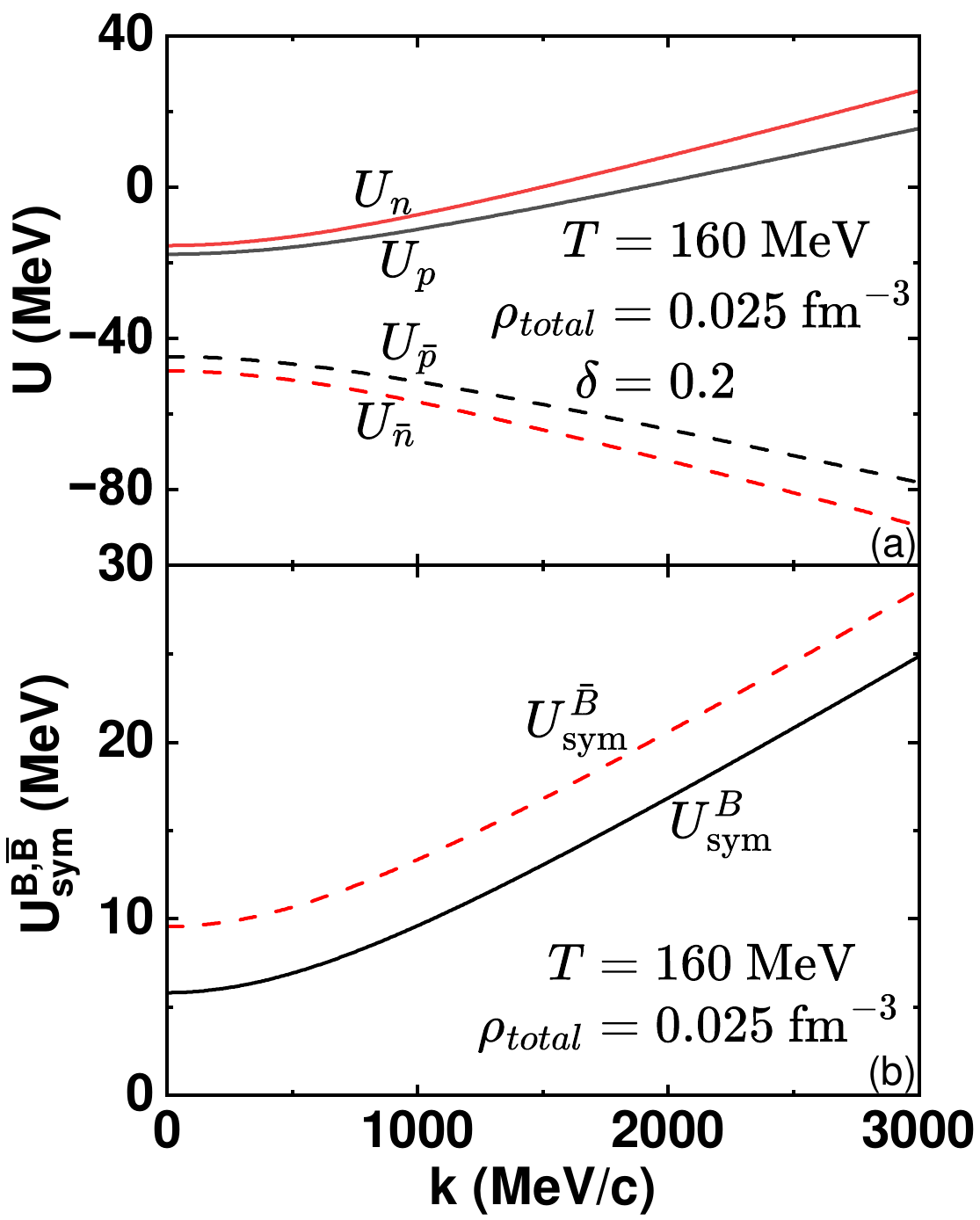}
\caption{\label{fig6} Upper: Neutron, proton, antineutron, and antiproton Schr\"odinger equivalent potential as a function of particle momentum in isospin-asymmetric baryon-rich nuclear matter at $T=160$ MeV, $\rho_{total}=0.025$ fm$^{-3}$, and $\delta=0.2$; Lower: Isospin symmetry potential of nucleons and antinucleons as a function of particle momentum in baryon-rich nuclear matter at $T=160$ MeV and $\rho_{total}=0.025$ fm$^{-3}$.}
\end{figure}

The isospin splitting of the nucleon and antinucleon Schr\"odinger equivalent potential is shown in Fig.~\ref{fig6}. In baryon- and neutron-rich nuclear matter at $T=160$ MeV, $\rho_{total}=0.025$ fm$^{-3}$, and $\delta=0.2$, the neutron potential is less attractive than the proton potential, while the antineutron potential is more attractive than the antiproton potential, as shown in Fig.~\ref{fig6} (a). We can defined the isospin symmetry potentials for both nucleons and antinucleons as
\begin{eqnarray}
U_{sym}^B &=& (U_n-U_p)/2\delta,\\
U_{sym}^{\bar{B}} &=& (U_{\bar{p}}-U_{\bar{n}})/2\delta,
\end{eqnarray}
and their momentum dependencies are shown in Fig.~\ref{fig6} (b). While both isospin symmetry potentials are stronger with increasing nucleon momentum, $U_{sym}^{\bar{B}}$ is seen to be stronger than $U_{sym}^B$, indicating a stronger isospin splitting of the mean-field potential for antinucleons compared to nucleons in baryon- and neutron-rich nuclear matter.

The difference in the isospin symmetry potentials of nucleons and antinucleons can be understood as follows. Considering the dominating terms $\Sigma_{sq}+\Sigma_{vq}$ in the Schr\"odinger equivalent potential [Eq.~(\ref{SEP})] at $k=0$, the isospin splittings of nucleon potential and antinucleon potential are
\begin{eqnarray}
U_n-U_p &\approx& 2 g_\delta \langle \delta \rangle - g_\rho \langle \rho_0 \rangle, \\
U_{\bar{p}} - U_{\bar{n}} &\approx& - 2 g_\delta \langle \delta \rangle - g_\rho \langle \rho_0 \rangle.
\end{eqnarray}
In the above, $\langle \delta \rangle$ and $\langle \rho_0 \rangle$ are negative in neutron-rich matter. The large $U_n$ than $U_p$ is due to the overwhelming contribution from $\rho$-meson than $\delta$-meson coupling. The larger $U_{\bar{p}} - U_{\bar{n}}$ is due to the additive contributions from $\rho$-meson and $\delta$-meson couplings, compared to the smaller $U_n-U_p$ due to the cancellation contributions from their couplings. In the absence of antinucleons, $\langle \rho_0 \rangle$ becomes more negative, which increases both $U_{sym}^B$ and $U_{sym}^{\bar{B}}$. At small $\rho_{\bar{B}}/\rho_{total}$, the existence of antinucleons may only affect slightly $U_{sym}^B$ and $U_{sym}^{\bar{B}}$ since they are linearly proportional to $\langle \rho_0 \rangle$, different from the case of the isospin symmetry energy shown in Fig.~\ref{fig5}.


To summarize, based on the relativistic mean-field model and assuming $G$-parity invariance, we have studied the equation of state of baryon- and neutron-rich nuclear matter, which can be formed in beam-energy-scan or fixed-target experiments at RHIC. Similar to the isospin symmetry energy, we have defined the baryon-antibaryon symmetry energy characterizing the energy difference due to the baryon-antibaryon asymmetry, and its potential contribution is correlated with the difference in the mean-field potential between nucleons and antinucleons in baryon-rich matter. At high temperatures, we found that the occurrence of even a small fraction of antinucleons reduces considerably the isospin symmetry energy, compared to the traditional case without antinucleons. In addition, it is observed that the antineutron potential is more attractive than the antiproton potential in baryon- and neutron-rich matter. Due to the additive contributions from $\rho$-meson and $\delta$-meson couplings, we found that the isospin symmetry potential for antinucleons is intrinsically larger than that for nucleons. Incorporating hyperons or nucleon resonances may increase the baryon density at chemical freeze-out in Fig.~\ref{fig1}, but will not change qualitatively the main conclusions obtained here. The present study is helpful in understanding properties of the produced baryon- and neutron-rich matter as well as its dynamics in low-energy relativistic heavy-ion collisions.

This work is supported by the National Natural Science Foundation of China under Grant Nos. 12375125 and 11922514, and the Fundamental Research Funds for the Central Universities.

\bibliography{EsymB}
\end{document}